\documentclass[conference]{IEEEtran}
\IEEEoverridecommandlockouts
\usepackage{cite}
\usepackage{amsmath,amssymb,amsfonts}
\usepackage{algorithmic}
\usepackage{graphicx}
\usepackage{textcomp}
\usepackage{xcolor}
\def\BibTeX{{\rm B\kern-.05em{\sc i\kern-.025em b}\kern-.08em
    T\kern-.1667em\lower.7ex\hbox{E}\kern-.125emX}}

\newtheorem{theorem}{Theorem}[section]

\newtheorem{lemma}[theorem]{Lemma}

\usepackage{algorithm}

\usepackage{tabularx}

\usepackage{url}

\usepackage{hyperref}

\hypersetup{
  citebordercolor=blue,
  filebordercolor=blue,
  linkbordercolor=blue,
  urlbordercolor=red
}

\begin{document}

\title{inGRASS: Incremental Graph Spectral Sparsification via Low-Resistance-Diameter Decomposition
}

\author{Ali Aghdaei \\Stevens Institute of Technology\\ aaghdae1@stevens.edu \and Zhuo Feng\\Stevens Institute of Technology\\zfeng12@stevens.edu}

\maketitle

\begin{abstract}
This work presents inGRASS, a novel algorithm designed for incremental spectral sparsification of large undirected graphs. The proposed inGRASS algorithm is highly scalable and parallel-friendly, having a nearly-linear time complexity for the setup phase and the ability to update the spectral sparsifier in $O(\log N)$ time for each incremental change made to the original graph with $N$ nodes. A key component in the setup phase of inGRASS  is a multilevel resistance embedding framework introduced for efficiently identifying spectrally-critical edges and effectively detecting redundant ones, which is achieved by decomposing the initial sparsifier into many node clusters with bounded effective-resistance diameters leveraging a low-resistance-diameter decomposition (LRD) scheme.  The update phase of inGRASS exploits low-dimensional node embedding vectors for efficiently estimating the importance and uniqueness of each newly added edge. As demonstrated through extensive experiments, inGRASS achieves up to over $200 \times$ speedups while retaining comparable solution quality in incremental spectral sparsification of graphs obtained from various datasets, such as circuit simulations, finite element analysis, and social networks.
\end{abstract}

% \begin{IEEEkeywords}
% spectral graph theory, incremental sparsification, effective resistance, graph decomposition
% \end{IEEEkeywords}

\section{Introduction}
The graph-based analysis is a crucial technique that finds extensive application in various electronic design automation (EDA) problems like logic synthesis and verification, layout optimization, static timing analysis (STA), network partitioning/decomposition, circuit modeling, and simulation. For instance, spectrally-sparsified graphs allow for accelerating circuit simulations, performing vectorless integrity verification of power grids, and analyzing worst-case on-chip temperature distributions.  

In recent years, mathematics and theoretical computer science (TCS) researchers have extensively studied various research problems related to simplifying large graphs using spectral graph theory \cite{ batson2012twice, spielman2011spectral,  Lee:2017}.
Specifically, recent research on spectral graph sparsification allows for the construction of much sparser subgraphs that can preserve important graph spectral (structural) properties like the first few eigenvalues and eigenvectors of the graph Laplacian. These findings have already led to the development of numerical and graph algorithms that can solve large sparse matrices and partial differential equations (PDEs) in nearly-linear time, as well as enable graph-based semi-supervised learning (SSL), computing the stationary distributions of Markov chains and personalized PageRank vectors, spectral graph partitioning and data clustering, max flow and multi-commodity flow of undirected graphs, and nearly-linear time circuit simulation and verification algorithms \cite{miller:2010focs, spielman2011spectral, feng2016spectral,zhuo:dac18}. 

However,  there still remain grand challenges when adopting spectral sparsification algorithms to real-world EDA applications:   existing spectral sparsification methods can not efficiently update the sparsified graph when only  \emph{incremental changes} are made to the original input graph \cite{feng2016spectral,feng2020grass,fegrass, zhang2020sf,liu2022pursuing}. For example, when a power grid network has been updated with a few additional metal wires connected to the system, traditional spectral sparsification methods (e.g. GRASS \cite{feng2020grass}, feGRASS \cite{fegrass}) must recompute the sparsifier from scratch, imposing a significant overhead during the chip optimization procedure. While dynamic algorithms for spectral graph sparsification have been recently studied to handle streaming edge insertions/deletions \cite{kapralov2020fast,filtser2021graph}, it remains unclear if such theoretical results would allow for practically efficient implementations. 
\begin{figure}
    \centering
\includegraphics [width = \linewidth]{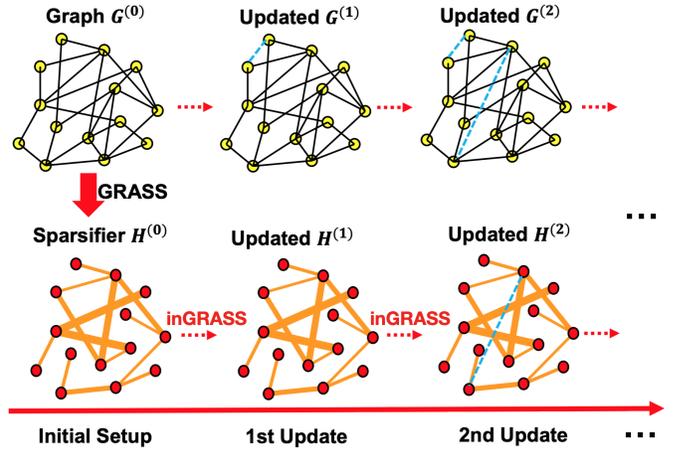}
    \caption{The proposed inGRASS  algorithm for incremental spectral   sparsification. Given the initial input graph $G^{(0)}$ and its sparsifier $H^{(0)}$, inGRASS constructs the updated sparsifiers $H^{(1)}, H^{(2)}, \cdots $ with newly added edges.}
    \label{fig:ingrass}
\end{figure}

To address the aforementioned limitations, this  work  presents a highly-efficient algorithmic framework (inGRASS) for incremental spectral graph sparsification (as shown in Fig. \ref{fig:ingrass}). A key component of inGRASS is an efficient effective-resistance embedding scheme in the setup phase, which leverages   low-resistance-diameter (LRD) decomposition   to partition the original sparsifier into multiple node clusters with bounded resistance diameters. We show that by exploiting a  multilevel LRD decomposition procedure, each node can be represented by a low-dimensional embedding vector that allows for extremely fast estimation of effective-resistance distances between any two  nodes  in the sparsifier. Motivated by recent spectral perturbation-based spectral sparsification methods \cite{feng2016spectral,feng2020grass,zhang2020sf} that prune each off-tree edge candidate based on its spectral distortions (defined as the product of the edge weight and effective resistance in the sparsifier), in the incremental update phase inGRASS  handles  each newly added edge by checking its spectral distortion that can be quickly estimated based on the proposed resistance embedding scheme,  which allows for identifying the most spectrally-critical edges while filtering out non-critical ones in the current sparsifier. Compared with re-running spectral sparsification algorithms from scratch, inGRASS has a much lower computational complexity for  incremental edge insertions:  for a weighted undirected graph with $O(N)$ nodes, the setup phase of inGRASS enjoys nearly-linear $O(N \log N)$ time  complexity as prior methods, while each incremental update of the sparsified graph can be accomplished in $O(\log N)$  time.   The key contribution of this work is summarized below:
\begin{enumerate}
\item  We propose an incremental spectral graph sparsification (inGRASS) algorithm for efficiently updating sparsified graphs considering newly added edges.
\item A key component of inGRASS is an effective-resistance embedding scheme that leverages multilevel LRD decomposition and allows for extremely fast effective-resistance calculations critical to spectral distortion analysis of each new edge inserted into the original graph. 
\item Our extensive experiment results on real-world VLSI designs show that inGRASS is over $200 \times$ faster than running state-of-the-art spectral sparsification algorithms from scratch while achieving comparable solution quality (condition number, graph density, etc).
% \item HyperEF has achieved very promising  results for various VLSI design datasets:   HyperEF always achieves the lowest conductance for most hypergraph coarsening tasks, while  preserving the original hypergraph cut as well as conductance.
\end{enumerate}

An implementation of our algorithm and the code for reproducing our experimental results are available online at  \url{https://github.com/Feng-Research/inGRASS}.

The rest of the paper is organized as follows. In Section  \ref{sec:background}, we provide a background introduction to the basic concepts related to spectral graph sparsification. In section \ref{sec:alg}, we introduce the  technical details of inGRASS as well as its algorithm flow and complexity.   In Section \ref{sec:result}, we demonstrate extensive experimental results to evaluate the performance of inGRASS using a variety of real-world VLSI design benchmarks, which is followed by the conclusion of this work in Section \ref{sec:conclusion}.

\section{Background}\label{sec:background}

\subsection{Spectral Graph Theory}
In a weighted undirected graph $G = (V, E, w)$, \(V\) (\(|V| = N\)) and \(E\) denote the sets of vertices and edges, \(w\) is a positive weight function, and \(w_{i,j}\) denotes the weight of the edge between vertices \(i\) and \(j\), or \(w(e)\) for edge \(e \in E\). The \(N \times N\) adjacency matrix \(A\) is symmetric (\(A(i,j) = A(j,i)\)) and positive semi-definite (\(x^{\top}A x \geq 0\) for any real vector \(x \in \mathcal{R}^{N\times 1}\)). The Laplacian matrix \(L_G\) is defined as \(L_G:= D - A\), where \(D\) is the degree matrix. Eigenvalues and eigenvectors of \(L_G\) reveal properties related to the graph structure, such as the number of connected components and node connectivity. The Laplacian quadratic form, \(x^{\top}L_G x\), is utilized in spectral graph theory to assess properties like graph cuts, clustering, and conductance, serving as an analytical tool for characterizing the Laplacian matrix \(L_G\) and the associated graph structure. It is also suggested that the spectral similarity between two graphs $G$ and $H$ can be measured using the following inequality \cite{spielman2008graph}:
\begin{equation}
\frac{x^{\top}L_G x}{\epsilon} \leq x^{\top}L_H x\leq \epsilon x^{\top}L_G x,
\end{equation}
where $L_G$ and $L_H$ are the Laplacian matrices of $G$ and $H$, respectively. Similarly, a smaller relative condition number $\kappa (L_G, L_H)$ implies a higher degree of spectral similarity between $G$ and $H$.

% Hence, the condition number $\kappa (G, H)$ serves as a quantitative measure to investigate and assess the level of spectral similarity between the graphs $G$ and $H$.

\subsection{Spectral Graph Decomposition}
\begin{lemma}\label{Lemma:Sc}
Spectral sparsification of an undirected graph $G = (V, E, w)$, with its Laplacian denoted by $L_G$, can be achieved by leveraging a short-cycle decomposition algorithm. This algorithm produces a sparsified graph $H = (V, E',w')$, where $E' << E$, with its Laplacian denoted by $L_H$, such that for all real vectors $x \in \mathcal{R}^{V}$, the approximation $x^\top L_G x \approx x^\top L_H x$ holds \cite{chu2020graph}.
\end{lemma}

Building upon Lemma \ref{Lemma:Sc}, the graph sparsification algorithm combines short-cycle decomposition with low-stretch spanning trees (LSSTs) \cite{abraham2012} to construct a sparsified graph that preserves the spectral properties of the original graph \cite{chu2020graph}.

Furthermore, a similar spectral graph sparsification method is proposed  in \cite{zhang2020sf}, which decomposes the graph into multiple sets of disjoint cycles using a multilevel spectral graph coarsening framework. By constructing a hierarchy of contracted graphs, the method identifies and adds spectrally-critical edges to the initial graph sparsifier.

\subsection{Effective Resistance}
For a weighted, undirected graph $G = (V, E, w)$ with $|V| = N$, the effective resistance between nodes $(p, q) \in V$ plays a crucial role in various graph analysis tasks including spectral sparsification algorithms \cite{spielman2008graph}. The effective resistance distances can be accurately computed using the equation:
\begin{equation}\label{eq:eff_resist0}
R_{eff}(p,q) = \sum\limits_{i= 2}^{N} \frac{(u_i^\top b_{pq})^2}{\lambda_i},
\end{equation}
where ${b_{p}} \in \mathbb{R}^{V}$ denote  the standard basis vector with all zero entries except for the $p$-th entry being $1$, and  $q$-th entry being $-1$  (${b_{pq}}=b_p-b_q$). $u_{i} \in \mathbb{R}^{\mathcal{V}}$ for $i=1,...,|V|$ denote the  unit-length, mutually-orthogonal  eigenvectors corresponding to  Laplacian eigenvalues $\lambda_i$ for $i=1,...,|V$.

\section{inGRASS: Incremental Graph Sparsification}\label{sec:alg}

In this section, we introduce inGRASS, an incremental graph spectral sparsification method in response to a stream of edge insertions, achieving nearly-linear time complexity. The key strength of inGRASS lies in its ability to rapidly identify spectrally-critical edges within the stream without the need for recomputing the edge importance metric for each new arrival. This feature significantly reduces the computational complexity of the algorithm, making inGRASS outperform existing graph sparsification and incremental edge update methods.

\subsection{Overview of inGRASS Algorithm}

The proposed inGRASS algorithm constructs a sparse data structure to enable fast estimation of the spectral distortion  and spectral similarity of  newly introduced edges. It consists of two primary phases: \emph{setup phase} and \emph{update phase}.

\paragraph{Setup Phase} 

This is a one-time operation that equips the inGRASS algorithm to iteratively update the initial graph sparsifier $H$ upon receiving streams of newly introduced edges in different iterations. In this phase, each node in the initial graph sparsifier $H^{(0)}$ is assigned a $\log N$-dimensional embedding vector by leveraging low-resistance-diameter (LRD) decomposition. A sparse data structure is then created that provides an efficient estimate of the resistance distance among different pair of nodes in the initial graph sparsifier $H^{(0)}$.

\paragraph{Update Phase}
Computing $O(\log N)$ node embedding vectors in the previous phase enables efficient spectral distortion estimation for newly added edges, identifying spectrally-critical edges. Subsequently, spectral similarity estimation is employed to enhance sparsification performance by selectively filtering edges with minimal impact on graph spectral properties. The determination of the filtering level is based on the target relative condition number and can be adjusted to achieve various degrees of spectral similarity between graph $G$ and graph $H$.

Combining these setup and update phases, inGRASS presents an efficient approach for incremental graph spectral sparsification, outperforming the existing methods.

\subsection{The Setup Phase of inGRASS}
\subsubsection{Scalable  Estimation of   Effective Resistances}
To address the computational complexity associated with directly computing eigenvalues and eigenvectors required for estimating edge  effective resistances, we introduce a scalable algorithm that approximates the eigenvectors of the graph Laplacian matrix using the Krylov subspace.
\begin{figure*}
    \centering
    \includegraphics [width=0.997588\textwidth]{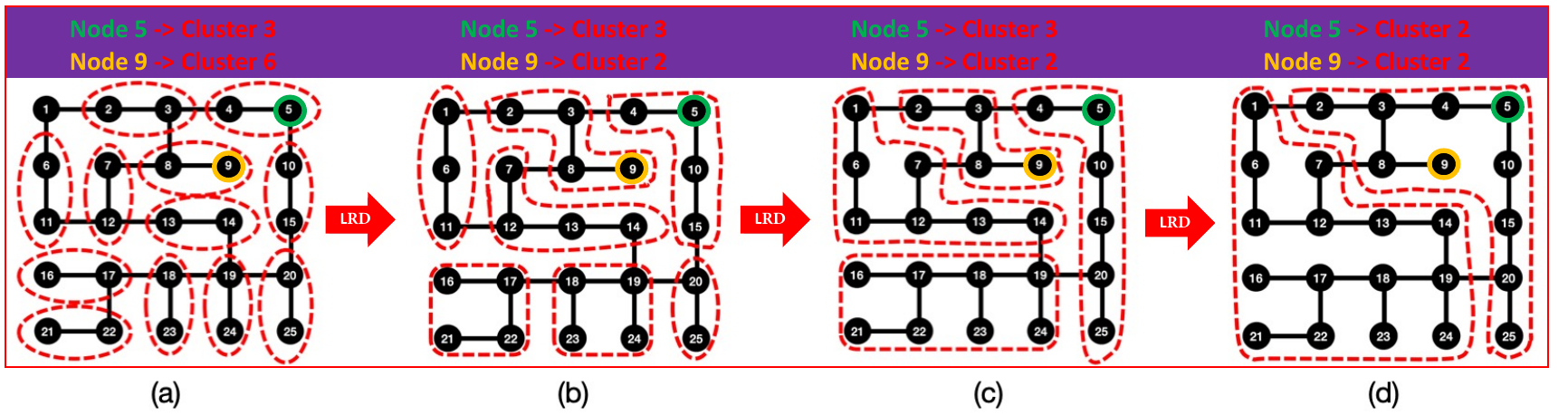}
    \caption{A 4-level resistance embedding of the initial graph sparsifier achieved through the proposed LRD decomposition. Since the embedding vectors for nodes $5$ and $9$ are $[3,3,3,2]^\top$ and $[6,2,2,2]^\top$,  the effective-resistance distance between them is bounded by the resistance diameter of cluster $2$ shown in (d). }
    \label{fig:LRD}
\end{figure*}

Let $A_{N \times N}$ denote the adjacency matrix of a graph, its Krylov subspace $\mathbf{K}_m(A, x)$ is a vector space spanned by the vectors computed through power iterations $x, A x, A^2 x, \ldots, A^{m-1} x$ \cite{liesen2013krylov}. By enforcing orthogonality among the above vectors in the Krylov subspace, we can compute a new set of mutually-orthogonal vectors of unit lengths for approximating the original Laplacian eigenvectors in   (\ref{eq:eff_resist0}), which are denoted as   $\Tilde{u}_{1}$, $\Tilde{u}_{2}$, $\ldots$, $\Tilde{u}_{m}$. To estimate the effective resistance between two nodes $p$ and $q$, we leverage the approximated eigenvectors of the Laplacian matrix $L_G$ constructed through the above Krylov subspace. Using   (\ref{eq:eff_resist0})  we obtain an estimation of each edge's effective resistance:
\begin{equation}\label{eq:ER_estimation}
R_{eff}(p,q) \approx \sum\limits_{i= 1}^{m} \frac{(\Tilde{u}^\top_{i}  b_{pq})^2}{\Tilde{u}^\top_{i} L_G \Tilde{u}_{i}},
\end{equation}
where  $\Tilde{u}_{i}$ represents the approximated eigenvector corresponding to the $i$-th eigenvalue of $L_G$, and $m$ is the order of the Krylov subspace.

\subsubsection{Multilevel Resistance Embedding }
We utilize a multilevel low-resistance-diameter (LRD) decomposition framework for embedding nodes of the initial graph sparsifier $H^{(0)}$ into an $O(\log N)$-dimensional space \cite{aghdaei2022hyperef}. The embedding process iteratively follows these steps: \textbf{(S1)} efficiently estimate the effective resistance of edges using (\ref{eq:ER_estimation}); \textbf{(S2)} contract edges starting from low-resistance-diameter clusters (cluster diameters are initialized to 0 for all nodes in the first level); \textbf{(S3)} replace a contracted edge with a supernode and adjust the  resistance diameter of the clusters accordingly.

% By examining the figure, we observe that increasing the $\beta$ value from left to right results in a smaller number of partitions and larger cluster sizes. This indicates that as the effective resistance diameter threshold is raised, nodes are grouped together into larger clusters, leading to a coarser level of graph decomposition. Conversely, lowering the $\beta$ value results in more partitions with smaller cluster sizes, representing a finer-grained decomposition.

\subsubsection{Multilevel Sparse Data Structure}
Leveraging the $O(\log N)$-dimensional node embedding vectors obtained in the previous step, we construct a specialized sparse data structure for efficient access to the node clustering indices at different levels. This facilitates efficient storage and retrieval of cluster connectivity information, enabling the identification of spectrally-critical and spectrally-unique edges among the newly added edges. The sparse data structure is promptly updated upon the addition of a newly introduced edge to the graph sparsifier.

Fig. \ref{fig:LRD} illustrates the computation of node embedding vectors through multilevel LRD decomposition of the initial graph sparsifier. Increasing the diameter thresholds, as shown in the figure, leads to larger cluster sizes progressing from left to right. This allows for estimating the upper bound of effective resistance between any two nodes based on the level where they share the same cluster index. For instance, nodes 5 and 9, with different cluster indices at levels (a), (b), and (c), share the same cluster index at level (d). Their resistance distance is bounded by the resistance diameter of cluster 2 at level (d).

\subsection{The Update Phase  of inGRASS}
\subsubsection{Spectral Distortion Estimation}
% \zf{Show how to compute the effective-resistance and spectral distortion (w*R) upper bound based on the aforementioned node embedding vectors}

It has been demonstrated that inserting a new edge with  a large spectral distortion \footnote{The spectral distortion of an  edge is defined as the product of its edge weight   and the effective resistance between its two end nodes \cite{feng2020grass,zhang2020sf}.} to graph sparsifier will significantly increase its first few Laplacian eigenvalues \cite{zhang2020sf, feng2020grass}.
\begin{lemma}
Let $H = (V, E', w')$, where $w':E' \rightarrow \mathbb{R}+$, denote the sparsified weighted graph of $G$, and $L_H$ denote its Laplacian matrix. The $i$-th Laplacian eigenvalue perturbation due to $\delta L_H=  w'_{p,q}b_{pq} b^\top_{pq}$ can be computed as:
\begin{equation}
\delta \lambda_{i} =  w'_{p,q} (u_{i}^{\top} b_{pq})^2,
\end{equation}
where $u_{i}$ represents the eigenvector corresponding to the $i$-th eigenvalue $\lambda_{i}$ of the Laplacian matrix $L_H$.
\end{lemma}
\begin{lemma}
Construct a  weighted eigensubspace matrix $U_K$ for $K$-dimensional spectral graph embedding using the first $K$  Laplacian eigenvectors and eigenvalues as follows:
\begin{equation}\label{subspace}
U_K=\left[\frac{u_2}{\sqrt {\lambda_2}},..., \frac{u_K}{\sqrt {\lambda_K}}\right], 
\end{equation}
then  the spectral distortion of the new edge $e_{p,q}$ will become the total  $K$-eigenvalue  perturbation $\Delta_K$ when $K \rightarrow N$ \cite{zhang2020sf}:
\begin{equation}\label{eq:delta}
\Delta_K=\sum\limits_{i = 2}^{{K}}  \frac{\delta { {\lambda}_{i}^{}}}{\lambda_i} =  w_{p,q} \|U_K^\top b_{pq}\|^2_2\approx w_{p,q} R(p,q).
\end{equation}
\end{lemma}

The equation (\ref{eq:delta}) proves that a higher spectral distortion $\Delta_K$ can be achieved by including edges with higher effective resistance $R(p,q)$ and higher edge weight $w_{p,q}$. Our algorithm efficiently estimates the effective resistance of the newly introduced edges, leveraging the node embedding information obtained from the setup phase. This enables sorting the newly introduced edges according to their estimated spectral distortion that leads to including the spectrally-critical edges to the graph sparsifier.

\subsubsection{Spectral Similarity Estimation}
The inGRASS algorithm applies an edge filtering process to exclude a newly added edge if there is already an existing edge in the graph sparsifier with a similar spectral distortion. This  filtering process minimizes the introduction of additional edges while still influencing the spectral distortion. The algorithm selects a filtering level ($\mathcal{L}$) from the levels computed during the LRD decomposition phase with respect to a target condition number. For a target condition number $\kappa (L_G,L_H) = C$, the filtering level ($\mathcal{L}$) is chosen with the maximum number of nodes in a cluster equal to $\frac{C}{2}$. This threshold bounds the maximum spectral distortion among nodes within clusters (equivalent to the shortest path among the nodes within each cluster) to achieve the desired condition number. 
\begin{figure}
    \centering
    \includegraphics [width = 0.99\linewidth]{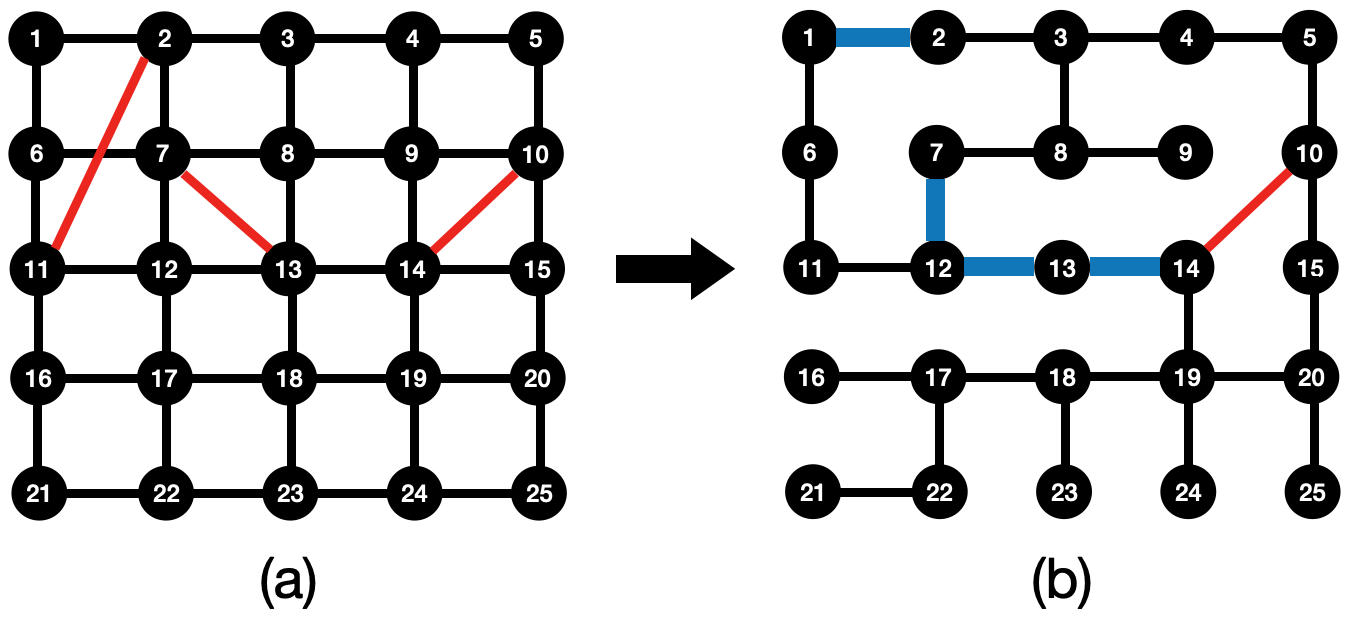}
    \caption{(a) The original graph featuring three newly introduced edges highlighted in red. (b) The edge included into the graph sparsifier, marked in red, alongside the edges with adjusted weights, denoted in blue.}
    \label{fig:update}
\end{figure}
For a newly introduced edge, inGRASS only includes the edge if there is no existing edge in the graph sparsifier connecting the clusters to which those two nodes belong in the specified filtering level ($\mathcal{L}$). In the event of an existing edge connecting the clusters of those nodes in the filtering level ($\mathcal{L}$), the newly introduced edge is discarded, and the weight of the existing edge is adjusted by adding the weight of the newly added edge to it. Moreover, if the newly introduced edge connects two nodes within the same cluster in the filtering level  ($\mathcal{L}$), the edge is discarded and its weight is proportionally distributed among the edges connecting the nodes within that cluster.

Fig. \ref{fig:update} illustrates the proposed edge filtering process employed by the inGRASS algorithm to select spectrally-critical and unique edges that will significantly impact the global structure of the sparsified graph. Let's consider three newly added edges to the original graph: $e_1 = (2,11)$, $e_2 = (7,13)$, and $e_3 = (10,14)$, as depicted in Fig. \ref{fig:update} (a), the algorithm determines the filtering level, denoted as $\mathcal{L}$ = (b) in Fig. \ref{fig:LRD}, to achieve a desired condition number of $\kappa (L_G,L_H) = 8$. Starting with $e_1 = (2,11)$, the algorithm integrates it into the existing edge $e = (1,2)$ between clusters 2 and 11, adjusting the weight accordingly. For $e_2 = (7,13)$, both nodes 7 and 13 share a cluster, leading to the exclusion of $e_2 = (7,13)$ and a proportional increase in weights of related edges. Finally, $e_3 = (10,14)$ is added as no existing edge connects the clusters of nodes 10 and 14 in the filtering level $\mathcal{L} = $ (b).

\subsection{The Algorithm Flow and Complexity Analysis}
The complete flow of the inGRASS algorithm is described in Algorithm \ref{alg:ingrass} to efficiently update the graph sparsifier  $H^{(0)} = (V, E', w')$ with $|V| = N$  after adding a set of weighted edges to the original graph $G^{(0)}$ to reach the target condition number $\kappa (L_G,L_H) = C$.

\begin{algorithm}
\small {\caption{inGRASS algorithm flow}\label{alg:ingrass}}
\textbf{Input:} $H^{(0)} = (V, E')$ where $|V| = N$, a set of  newly introduced edges to $G^{(0)}$, a target condition number $\kappa (L_G,L_H) = C$.\\
\textbf{Output:} The updated graph sparsifier $H$.\\
  \algsetup{indent=1em, linenosize=\small} \algsetup{indent=1em}
    \begin{algorithmic}[1]  
    \STATE \textbf{Setup Phase 1:} Effective resistance estimation using (\ref{eq:ER_estimation}).
    \STATE \textbf{Setup Phase 2:} Node embeddings via LRD decomposition.
    \STATE \textbf{Setup Phase 3:} $O(\log N)$-level sparse data structure.
    \STATE \textbf{Update Phase 1:} Spectral distortion estimation of new edges.
     \STATE \textbf{Update Phase 2:} Spectral similarity estimation according to $C$.
    \STATE Return $H$.
    \end{algorithmic}
\end{algorithm}

The complexity analysis of the inGRASS algorithm is split into two phases:
\begin{itemize}
\item In the setup phase, the $O(\log N)$-level multilevel low-resistance-diameter (LRD) decomposition is applied to compute node cluster indices, resulting in a complexity of $O(N\log N)$.
\item In the update phase, the complexity arises from calculating the spectral distortion and checking spectral similarity of newly added edges using the $O(\log N)$-dimensional node embedding vectors from the setup phase, with a complexity of $O(\log N)$ for each new edge in the original graph.
\end{itemize}

\section{Experimental results}\label{sec:result}
\begin{table}[]
\caption{GRASS time vs inGRASS setup time.}
\centering
\begin{tabular}{ccccc}
\hline
Test Cases          & $|V|$    & $|E|$    & GRASS (s) & Setup (s) \\ \hline
G3 circuit    & 1.5E+6 & 3.0E+6 & 18.7 s    & 13.7  s    \\
G2 circuit    & 1.5E+5 & 2.9E+5 & 0.75 s    & 0.9    s   \\
fe\_4elt2     & 1.1E+4 & 3.3E+4 & 0.053 s    & 0.06    s  \\
fe\_ocean     & 1.4E+5 & 4.1E+5 & 1.12 s    & 1.01     s \\
fe\_sphere    & 1.6E+4 & 4.9E+4 & 0.08 s     & 0.17     s \\
delaunay\_n18 & 2.6E+5 & 6.5E+5 & 2.2 s     & 1.9      s \\
delaunay\_n19 & 5.2E+5 & 1.6E+6 & 6.2 s     & 4        s \\
delaunay\_n20 & 1.0E+6 & 3.1E+6 & 14.1 s    & 9.5      s \\
delaunay\_n21 & 2.1E+6 & 6.3E+6 & 28.5  s    & 19  s      \\
delaunay\_n22 & 4.2E+6 & 1.3E+7 & 62 s      & 38.6  s    \\
M6            & 3.5E+6 & 1.1E+7 & 83 s      & 45     s   \\
333SP         & 3.7E+6 & 1.1E+7 & 84 s      & 46      s  \\
AS365         & 3.8E+6 & 1.1E+7 & 84 s      & 48      s  \\
NACA15        & 1.0E+6 & 3.1E+6 & 13.8 s    & 8       s \\ \hline
\end{tabular}
\label{tab:case infos}
\end{table}

This section presents the results of a diverse range of experiments conducted to assess the performance and efficiency of the proposed incremental graph spectral sparsification algorithm (inGRASS). The test cases were selected from a wide array of matrices commonly utilized in circuit simulation, and finite element analysis applications \footnote{https://sparse.tamu.edu/}. All experiments were conducted on a Linux Ubuntu system with 1 terabyte of RAM and a $3.6$ GHz $64$-core CPU. The state-of-the-art spectral sparsification tool GRASS \cite{feng2020grass} \footnote{https://sites.google.com/mtu.edu/zhuofeng-graphspar/home} has been used as the benchmark to evaluate the performance and scalability of inGRASS. 

Table \ref{tab:case infos} provides a runtime comparison between GRASS and inGRASS setup times (measured in seconds) for various test cases. The runtime for the setup phase of inGRASS is mainly due to the LRD decomposition step for computing resistance embeddings for the initial graph sparsifier $H^{(0)}$. Note that this step is a one-time task, which can be leveraged for many subsequent update iterations involving streaming edge modifications. We observe that the inGRASS setup time is even faster than the GRASS time for most test cases as shown in   Table \ref{tab:case infos}, where $|V|$ ($|E|$) denotes the number of nodes (edges) in the original graph $G^{(0)}$, respectively.
\subsection{inGRASS for Incremental Spectral Sparsification}
\begin{table*}[]
\caption{Comparison of Incremental Graph Spectral Sparsification Outcomes through $10$-Iterative Updates using GRASS, inGRASS, and Random Methods.}
\centering
\begin{tabular}{|c|c|c|c|c|c|c|c|c|}
\hline
Test   Cases  & Density (D)                          & $\kappa (L_G,L_H)$        & GRASS-D & inGRASS-D & Random-D & GRASS-T & inGRASS-T & $\frac{GRASS-T}{inGRASS-T}$ \\ \hline
G3 circuit    & 10.0\% $\rightarrow$ 34\%   & 88 $\rightarrow$ 353  & 11.6\% & 11.7\% & 23.0\%   & 196 s & 1.7  s & 115 $\times$ \\ \hline
G2 circuit    & 9.0\% $\rightarrow$  32\%   & 72 $\rightarrow$ 283   & 11.0\% & 11.4\% & 25.7\%   & 7.8  s& 0.11  s& 71 $\times$\\ \hline
fe\_4elt2     & 10.0\%  $\rightarrow$ 39\%  & 95 $\rightarrow$ 330   & 10.0\% & 10.1\% & 36.0\%   & 0.56 s& 0.008 s& 70  $\times$\\ \hline
fe\_ocean     & 9.8\%   $\rightarrow$ 50\%  & 210 $\rightarrow$ 468   & 9.6\%  & 11.5\% & 30.3\%   & 11.8 s& 0.13  s& 91  $\times$\\ \hline
fe\_sphere    & 10.5\%  $\rightarrow$ 41\%  & 123 $\rightarrow$ 1103  & 8.2\%  & 10.5\% & 34.9\% & 0.84 s& 0.009 s & 93  $\times$\\ \hline
delaunay\_n18 & 10.5\%  $\rightarrow$ 35\%  & 113 $\rightarrow$ 336   & 11.7\% & 11.6\% & 30.5\%   & 23.1 s& 0.19  s& 122 $\times$\\ \hline
delaunay\_n19 & 10.6\%  $\rightarrow$ 35\% & 122 $\rightarrow$ 406 & 11.9\%  & 11.8\%    & 29.7\%   & 65.1   s & 0.41 s     & 159 $\times$                        \\ \hline
delaunay\_n20 & 10.5\%  $\rightarrow$  35\% & 126 $\rightarrow$ 418  & 11.7\% & 11.7\% & 30.2\%   & 148 s & 0.9  s & 164 $\times$\\ \hline
delaunay\_n21 & 10.0\%  $\rightarrow$  35\% & 151 $\rightarrow$ 428  & 11.0\% & 10.9\% & 29.5\%   & 299  s& 2.1 s  & 142 $\times$\\ \hline
delaunay\_n22 & 10.3\%  $\rightarrow$  34\% & 150 $\rightarrow$ 491  & 10.7\% & 10.7\% & 30.7\%   & 651 s & 4.3  s & 151 $\times$\\ \hline
M6            & 9.8\%  $\rightarrow$   34\% & 172 $\rightarrow$ 817   & 10.2\% & 11.2\% & 29.3\%   & 871  s& 4  s   & 218 $\times$\\ \hline
333SP         & 9.7\%  $\rightarrow$   34\% & 180 $\rightarrow$ 897  & 9.5\%  & 11.0\% & 29.4\%   & 882 s & 4.2 s  & 210 $\times$\\ \hline
AS365         & 10.1\% $\rightarrow$   34\% & 157 $\rightarrow$ 1876 & 10.4\% & 12.7\% & 31.1\%   & 885  s& 4.5  s & 197 $\times$\\ \hline
NACA15          & 10.4\% $\rightarrow$  34\%  & 152 $\rightarrow$ 585  & 10.4\% & 11.5\% & 29.3\%   & 145 s & 1  s   & 145 $\times$\\ \hline
\end{tabular}
\label{tab:all cases}
\end{table*}
Table \ref{tab:all cases} presents a thorough comparison of our incremental graph spectral sparsification algorithm, inGRASS, with GRASS and Random methods through $10$-iterative updates. The density ($D$), defined as $D:= \frac{|E|}{|V|}$, shows the density of the initial graph sparsifier $H^{(0)}$ and its density when all newly introduced edges are included. This emphasizes the importance of incremental spectral sparsification to prevent a substantial increase in graph sparsifier density. We maintain an initial density of $D = 10\%$ for consistency. The relative condition number $\kappa (L_G,L_H)$ measures the spectral similarity between the original graph $G$ and the updated graph sparsifier $H$. A smaller relative condition number indicates higher spectral similarity. The column related to $\kappa (L_G,L_H)$ indicates how the condition number between $G^{(0)}$ and $H^{(0)}$ is perturbed when newly added edges are excluded from $H^{(0)}$, providing insights into edge generation. Columns "GRASS-D," "inGRASS-D," and "Random-D" compare  graph sparsifier density for a target condition number using each method. \textbf{The target condition number is set to match the initial condition number between $G^{(0)}$ and $H^{(0)}$.} For instance, the target condition number for "G3 circuit" is set to $\kappa (L_{G^{(0)}}, L_{H^{(0)}}) = 88$. Notably, inGRASS achieves the target condition number comparably to GRASS and significantly outperforms Random, accompanied by a runtime speedup exceeding $\frac{GRASS-T}{inGRASS-T} = 200 \times$ across all test cases through $10$-iterative updates.

\subsection{ Robustness of the inGRASS Algorithm}
The performance of our incremental graph spectral sparsification algorithm, inGRASS, is studied for the "G2 circuit" dataset over different initial graph sparsifier densities. In Table \ref{tab:G2 case}, the labeled column "Density (D)" shows an initial density range of $6.5\%$ to $12.7\%$, with the graph sparsifier density set to $32\%$ when all edges are included. The $\kappa (L_G,L_H)$ column indicates the perturbation in the condition number between the original graph $G^{(0)}$ and the initial graph sparsifier $H^{(0)}$ when no newly added edges are included. The target condition number is set to the initial value (e.g. 56 for $D = 12.7\%$). Both "GRASS-D" and "inGRASS-D" denote the graph sparsifier density for the same target condition number, with inGRASS showing comparable results to GRASS across various initial densities.
\begin{table}[]
\caption{GRASS vs inGRASS densities across different initial densities in the graph sparsifier (``G2 circuit" test case).}
\centering
\begin{tabular}{|c|c|c|c|}
\hline
Density (D)                & $\kappa (L_G,L_H)$        & GRASS-D & inGRASS-D \\ \hline
12.7\% $\rightarrow$ 32\%  & 56 $\rightarrow$ 190  & 13.1\% & 14.8\%   \\ \hline
11.8\% $\rightarrow$  32\% & 66 $\rightarrow$ 196  & 12.0\%    & 13.8\%   \\ \hline
9.0\%  $\rightarrow$ 32\%  & 72 $\rightarrow$ 283  & 11.0\%    & 11.4\%   \\ \hline
7.6\%   $\rightarrow$ 32\% & 87 $\rightarrow$ 373  & 8.6\%  & 8.8\%    \\ \hline
6.6\%  $\rightarrow$ 32\%  & 103 $\rightarrow$ 432 & 7.7\%  & 7.9\%    \\ \hline
\end{tabular}
\label{tab:G2 case}
\end{table}
\subsection{Runtime Scalability of   inGRASS  }
\begin{figure}
    \centering
    \includegraphics [width = 0.9\linewidth]{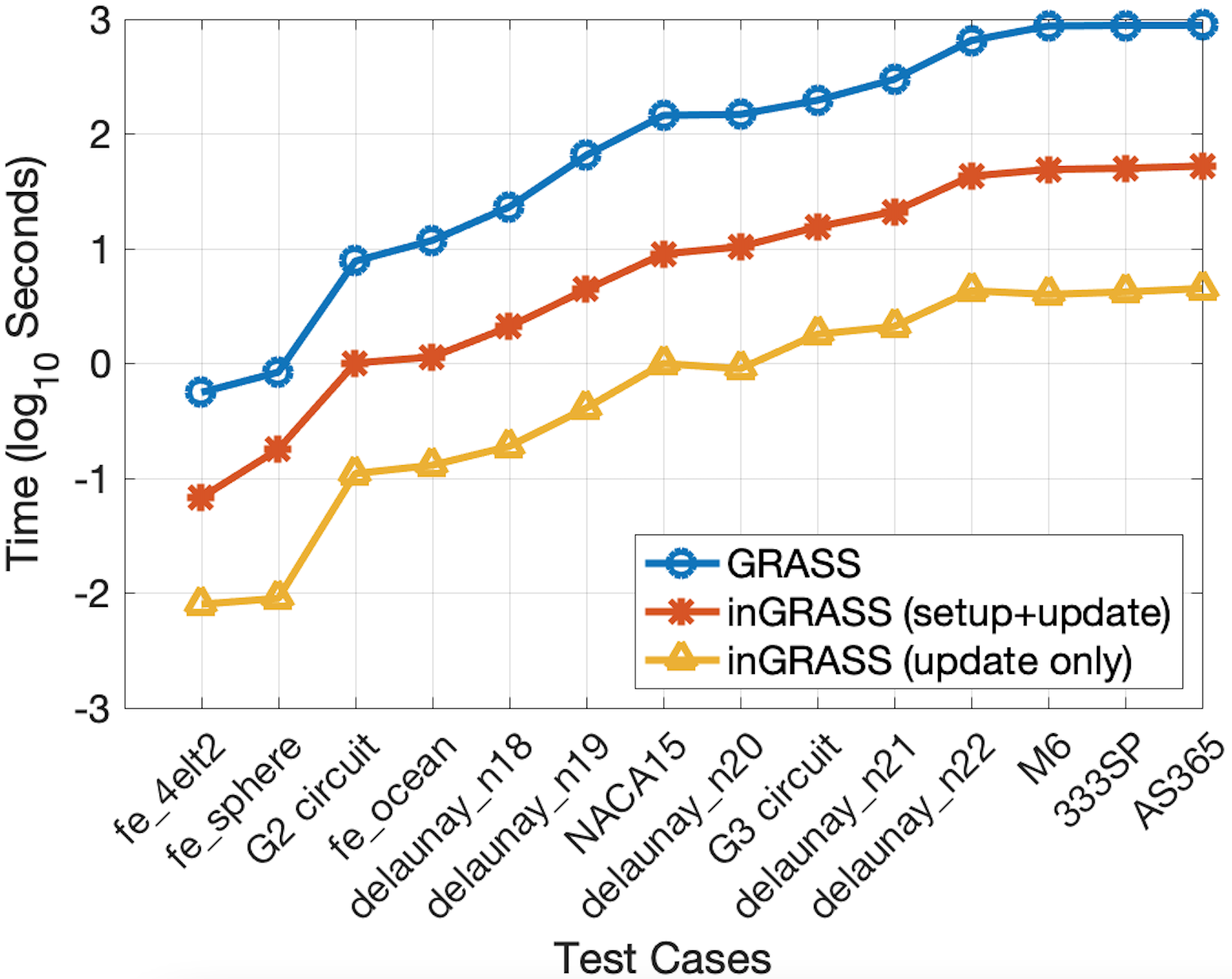}
    \caption{Runtime scalability comparison between GRASS and inGRASS.}
    \label{fig:runtime}
    \vspace{-20pt}
\end{figure}
Fig. \ref{fig:runtime} illustrates the runtime scalability of inGRASS compared with GRASS on a logarithmic scale for various test cases. It is demonstrated that inGRASS is faster than GRASS by over $200 \times$ during $10$-iterative updates. The runtime of inGRASS is also shown by adding the one-time setup time to provide a more comprehensive insight into the overall runtime of inGRASS compared with GRASS. %Additionally, the substantial speedup observed in inGRASS highlights its efficiency and suitability for applications requiring rapid graph spectral sparsification.

\section{Conclusion}\label{sec:conclusion}
This work presents inGRASS, a  highly scalable and parallel-friendly algorithm for incremental spectral sparsification of large undirected graphs.  The algorithm leverages a low-resistance-diameter decomposition (LRD) scheme to decompose the sparsifier into small clusters with bounded effective-resistance diameters. A multilevel resistance embedding framework is introduced for efficiently identifying spectrally-critical edges as well as detecting redundant ones. The proposed inGRASS achieves state-of-the-art results for incremental spectral sparsification of various networks derived from circuit simulations, finite element analysis, and social networks.

\section*{Acknowledgment}
This work is supported in part by the National Science Foundation under Grants CCF-2205572, CCF-2021309, CCF-2212370, and CCF-2011412.

\bibliographystyle{IEEEtran}
\bibliography{Ref,feng}

\end{document}